\documentclass[12pt]{article}
\usepackage{amsmath,amsbsy,amssymb,graphics}
\usepackage{graphicx,epsfig}
\def\beq{\begin{equation}}
\def\eeq{\end{equation}}
\def\be{\begin{equation}}
\def\ee{\end{equation}}
\def\bea{\begin{eqnarray}}
\def\eea{\end{eqnarray}}

\newcommand{\gsim}{\lower.7ex\hbox{$\;\stackrel{\textstyle>}{\sim}\;$}}
\newcommand{\lsim}{\lower.7ex\hbox{$\;\stackrel{\textstyle<}{\sim}\;$}}

\begin{document}

\begin{center}
 \vspace{0.2cm}
 {\Large \bf Emission angle distribution and flavor transformation of supernova neutrinos}

\vspace{0.6cm} {\large \bf Wei Liao}

\vspace{0.3cm} {

 Institute of Modern Physics,
 East China University of Science and Technology, \\
 P.O. Box 532, 130 Meilong Road, Shanghai 200237, P.R. China

 \vskip 0.3cm

 Center for High Energy Physics, Peking University, Beijing 100871, P. R. China

}
\end{center}

\begin{abstract}
 \vskip 0.2cm
 Using moment equations we analyze collective flavor transformation
 of supernova neutrinos. We study the convergence of moment equations
 and find that numerical results using a few moment converge quite
 fast. We study effects of emission angle distribution of neutrinos
 on neutrino sphere. We study scaling law of the amplitude of neutrino
 self-interaction Hamiltonian and find that it depends on model of
 emission angle distribution of neutrinos. Dependence of neutrino
 oscillation on different models of emission angle distribution is
 studied.

\end{abstract}


\section{Introduction}\label{sec1}
 Flavor transformation of neutrinos in core-collapse supernova is one of the
 important remaining problems in neutrino physics.
 This problem has been investigated by many researchers ~\cite{Pan,saw,
 dfq0,dfcq1,hrsw,dfc,rs1,dfcz,flmm,flmmt,dfcq2,dfq,rs2,dd,
 dfq3,ddmr,ccdk,gv,sf,dfcq4,prs,RS,Liao}. It is realized that neutrino
 density above neutrino sphere in supernova can be so large that
 neutrino-neutrino refraction can dominate flavor transformation of neutrinos.
 Research on the effect of neutrino-neutrino refraction is difficult because
 it is caused by neutrino self-interaction and is of non-linear nature.
 Complete numerical analysis use discrete set of energy bins and
 angle bins of neutrinos. Evolutions of million equations have to be followed.
 It is very complicated.

 In a recent work we derived a set of moment equations describing the
 transport and flavor transformation of neutrinos in supernova~\cite{Liao}.
 Distribution of neutrinos over angle $\theta_p$, the angle of
 neutrino direction intersecting with radial direction in supernova, is
 encoded in moments of density matrix. The equation of neutrino is
 expanded using these moments. Instead of using a large number of
 angle bins we just need a few moments to do numerical study. It is shown
 that numerical works can be simplified by about two orders of magnitude
 in comparison with multi-angle simulation. Moreover, this formulation
 of neutrino in supernova also offers us a way to study the
 effect of emission angle distribution of neutrinos on the transport
 and flavor transformation of neutrinos.

 In this article we analyze the effect of emission angle distribution
 of neutrinos. In section \ref{sec2} we make a quick review on the moment
 equations. In section \ref{sec3} we analyze the scaling behavior of
 the strengths of moments in different models of neutrino emission.
 We check convergence property of moment equations in the analysis of
 the strengths of moments. In section \ref{sec4} we analyze effect
 of different models of neutrino emission on collective neutrino oscillation.
 We summarize in section \ref{sec5}.

\section{Moment equations} \label{sec2}
\begin{figure}
\begin{center}
\includegraphics[height=5.cm,width=7.8cm]{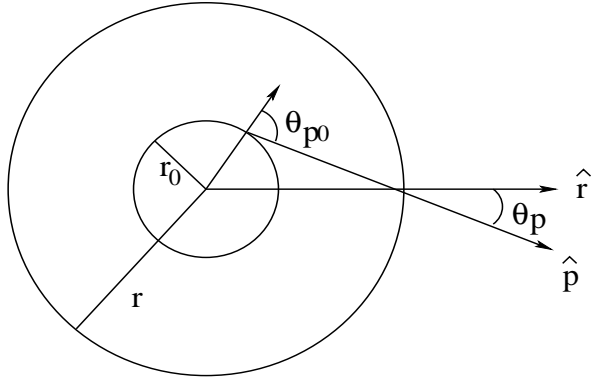}
\end{center}
 \vskip 0.0cm
  \caption{\small Geometric picture of angles of the neutrino momentum
  intersect with ${\hat r}$.}
 \label{fig1}
\end{figure}

 In Ref. ~\cite{Liao} we introduced moments of $\rho_{\vec p}(t,r)$, density matrix for
 neutrinos at given time $t$ and radius $r$:
 \bea
 \rho_k(t,r,|{\vec p}|)= \int d\Omega_{\vec p} ~
 (1-\cos\theta_p)^k ~\rho_{\vec p}(t,r),~k=0,1,2\cdots, \label{rhok}
 \eea
 where $\theta_p$ is the angle of neutrino direction intersecting with the
 radial direction, as shown in Fig \ref{fig1}.
 Similarly we introduced ${\bar \rho}_k$ for anti-neutrinos.
 We also introduced re-scaled moments
 \bea
 \rho^\prime_k=z^{2(k+1)} \rho_k, \label{rhokp}
 \eea
 where
 \bea
 z=r/r_0,
 \eea
 $r_0$ is the radius of neutrino sphere.
 Similarly we introduced ${\bar \rho}^\prime_k$ for anti-neutrinos.
 In Fig. \ref{fig1} one can see clearly
 \bea
 \sin\theta_p= \frac{r_0}{r} \sin\theta_{p0} \label{geom1}
 \eea
 It is easy to see
 \bea
 1-\cos\theta_p=\bigg( r_0^2/r^2 \bigg)\bigg/ \bigg(1+\sqrt{1-\frac{r^2_0}{r^2} \sin^2
 \theta_{p0} } ~\bigg). \label{geom2}
 \eea
 It scales approximately as $r^{-2}$.
 Together with the scaling behavior of the zeroth moment
 a geometric factor $z^{-2(k+1)}$ is found for moment
 $\rho_k$. The factor $z^{2(k+1)}$ is introduced in Eq. (\ref{rhokp})
 to compensate this geometric scaling factor.

 Using some approximations we arrive at the following
 set of moment equations
  \bea
 \frac{d \rho^\prime_k}{d r}&& =- r_0^{-1} Q^1_k-i[H_A,\rho^\prime_k],
 ~~k=0,1,\cdots,N \label{LVEq}
 \eea
 where $N\geq 1$ is an integer
 \bea
 && Q^1_k= z^{2k} \sum_{l=k+1}^{N}(l+1)z^{-(2l+ 1)} \rho^\prime_l,  \label{LVEqa}\\
 && H_A=H_0+\sqrt{2}G_F(L+z^{-4} D_1), \label{LVEqb} \\
 && D_1=\int \frac{dE}{(2 \pi)^3} ~E^2
 ~[\rho^\prime_1(r,E)-{\bar \rho}^\prime_1(r,E)]. \label{LVEqc}
 \eea
 Eq. (\ref{LVEq}) is a set of truncated moment equations in $P_N$
 approximation
 for which $\rho^\prime_k=0$ (${\bar \rho}^\prime_k=0$) has been set for $k>N$.
 $Q^1_N=0$. $H_0$ is the Hamiltonian for vacuum oscillation,
 $L=diag\{n_e,n_{\mu},n_{\tau}\}$ in the flavor base is the matter
 term given by charged lepton number densities $n_{e,\mu,\tau}$.
 $G_F$ is the Fermi constant. Equation for ${\bar \rho}_k$ is similar
 except replacing $H_0$ by $-H_0$.

 A few points concerning moment equations
 are as follows:
 a) Physical observables are described by $\rho_0$ and $\rho_1$.
 Integration of $E^2 Tr[\rho_0]$ over energy gives the neutrino density and
 integration of $E^2 Tr[\rho_0-\rho_1]$ gives the
 neutrino flux; b) Emission angle distribution of neutrinos on neutrino sphere is
 described by moments $\rho_k$ and their effect in the neutrino
 flavor transformation can be systematically studied;
 c) The strength of $\rho^\prime_k$, $Tr[\rho^\prime_k]$, is modified by
 $Q^1_k$ term and does not change if this term is neglected;
 d) The scaling law of the self-interaction Hamiltonian is no longer
 $z^{-4}$ when $N >1$ and is modified by higher moments. Precise
 scaling behavior should depend on the model of neutrino emission.

 \section{Emission angle distribution and scaling law of moments}\label{sec3}
 In this section we analyze the strengths of zeroth and first moments,
 that is $Tr[\rho_{0,1}]$, in different models of neutrino emission. This
 analysis can tell us a lot on how strong neutrino self-interaction is.
 It can also tell us a lot on
 the convergence property of moment equations. This is because $\rho_0$
 and $\rho_1$ are the most important
 quantities in our problem. Physical observables are given by
 $\rho_0$ and $\rho_1$. When neutrino self-interaction gives dominant
 contribution flavor transformation of neutrinos is controlled by $D_1$ which is
 directly related to $\rho_1$. Effects of higher moments on
 $Tr[\rho_{0,1}]$ tell us how large higher moments
 affect the flavor transformation of neutrinos.

      \begin{figure}
\begin{center}
\includegraphics[height=7.cm,width=16cm]{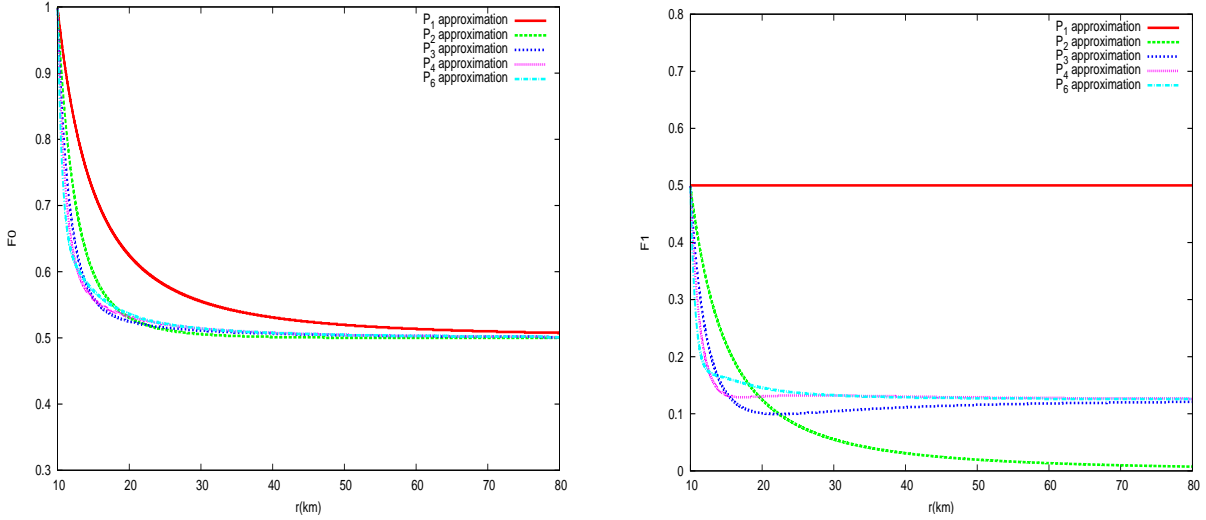}
\end{center}
 \vskip 0.0cm
  \caption{\small (color online) $F_0$ and $F_1$ in Model I.
   }
 \label{fig2}
\end{figure}

  We consider three models of neutrino emission on neutrino sphere.

 Model I, neutrino is uniformly emitted with
 respect to the emission angle $\theta_{p0}$ and
 \bea
 \rho_k(t, r_0)= \frac{1}{k+1} \rho_0(t, r_0) \label{model1}
 \eea

 Model II, emission angle distribution of neutrinos is
 proportional to $\cos\theta_{p0}$ and
 \bea
 \rho_k(t,r_0)=\frac{2}{(k+1)(k+2)}~\rho_0(t,r_0), \label{model2}
 \eea

 Model III, emission angle distribution of neutrinos is
 proportional to $(1-\cos\theta_{p0})\cos\theta_{p0}$ and
 \bea
  \rho_k(t,r_0)=\frac{6}{(k+2)(k+3)}~\rho_0(t,r_0). \label{model3}
 \eea

 The evolution of $Tr[\rho_k]$ is simple and is obtained by taking
 the trace of Eq. (\ref{LVEq}):
 \bea
 \frac{d Tr[\rho^\prime_k]}{d r}&& =- r_0^{-1} Q_k \label{LVEq1}
 \eea
 where
 \bea
 Q_k= z^{2k}
 \sum_{l=k+1}^{N}(l+1)z^{-(2l+ 1)} Tr[\rho^\prime_l].
 \label{LVEq1a}
 \eea
 $Q_N=0$. The second term in (\ref{LVEq}) does not contribute
 to $Tr[\rho^\prime_k]$.

     \begin{figure}
\begin{center}
\includegraphics[height=7.cm,width=16cm]{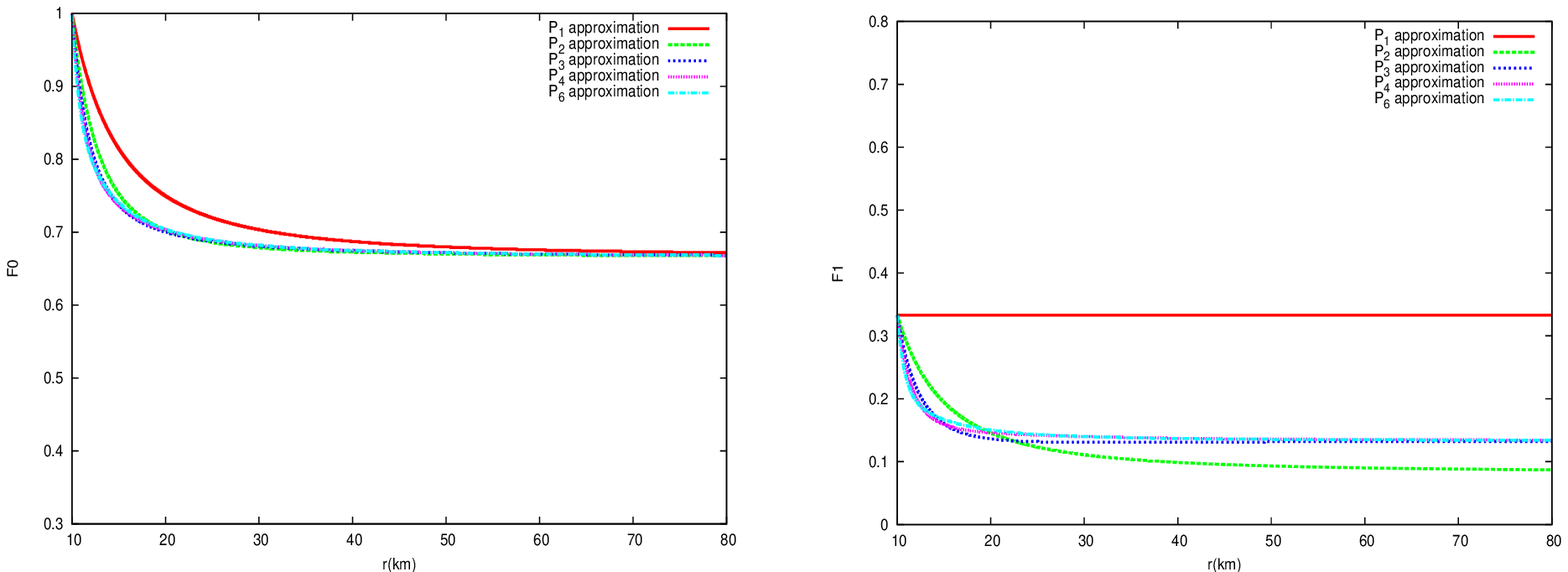}
\end{center}
 \vskip 0.0cm
  \caption{\small (color online) $F_0$ and $F_1$ in Model II.
   }
 \label{fig3}
\end{figure}

     \begin{figure}
\begin{center}
\includegraphics[height=7.cm,width=16cm]{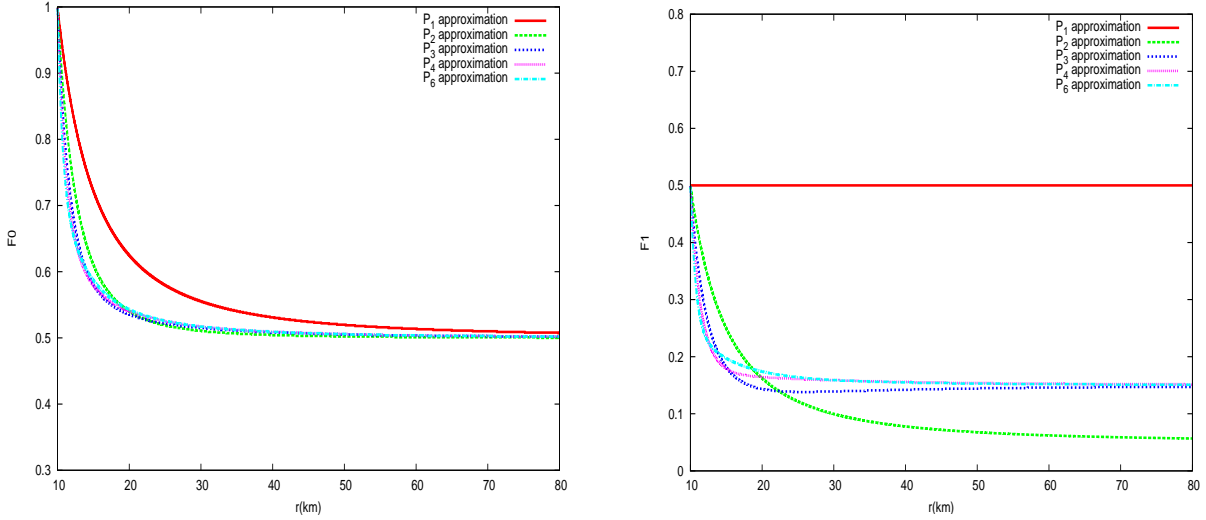}
\end{center}
 \vskip 0.0cm
  \caption{\small (color online) $F_0$ and $F_1$ in Model III.
   }
 \label{fig4}
\end{figure}

 We do numerical analysis for $F_0$ and $F_1$:
 \bea
 F_0=Tr[\rho'_0(r)]/Tr[\rho'_0(r_0)],~~
 F_1=Tr[\rho'_1(r)]/Tr[\rho'_0(r_0)] \label{ratio}
 \eea
 $F_{0,1}$ are $Tr[\rho'_{0,1}]$ relative to
 $Tr[\rho'_0]$ at $r=r_0$.
 In our numerical analysis we work in two flavor system of $(\nu_e, \nu_x)$.
 We choose $L_{\nu_e}=L_{{\bar \nu}_e}=L_{\nu_x}=L_{{\bar \nu}_x}
  =3.\times 10^{51}$ erg$/$s.
  The initial energy spectrum of neutrino is given by the Fermi-Dirac
  distribution
  \bea
  f_\nu(E)=\frac{1}{N_2~T_\nu}
  \frac{x^2}{e^{x-\mu_\nu}+1}, \label{dist}
  \eea
  where $x=E/T_\nu$ and $N_2$ is the normalization factor.
  Parameters of four types of neutrinos and anti-neutrinos are chosen as:
  $T_{\nu_e}=2.76$ MeV, $T_{{\bar \nu}_e}=4.01$ MeV,
  $T_{\nu_x}=T_{{\bar \nu}_\mu}=6.26$ MeV.
  $\mu_{\nu_e}=\mu_{{\bar \nu}_e}=\mu_{\nu_x}=\mu_{{\bar
 \nu}_x} =3.$

  In Fig. \ref{fig2}, \ref{fig3} and \ref{fig4} we show results in Model I,
  II and III separetely. A number of characteristics can be read out in these figures:

  i) In $P_1$ approximation $F_1$ keeps as a constant. This
  is because $Q_1$ is set to zero in this approximation.

  ii) In $P_2$ approximation $F_1$ is modified. Results of $F_0$ do
  not agree with those in $P_1$ approximation.

  iii) In $P_3$ approximation results of $F_0$ become close to those in $P_2$
  approximations. Results of $F_1$ do not agree with those in
  $P_2$ approximation. This is because in $P_2$ approximation
  $F_1$ becomes corrected by $Tr[\rho^\prime_2]$ but $Tr[\rho^\prime_2]$
  is still a constant. In $P_3$ approximation $Tr[\rho^\prime_2]$ is
  also corrected and its contribution to $F_1$ is modified.

  iv) Results of Model II and III in $P_4$ and $P_6$ approximations
  agree perfectly for both $F_0$ and $F_1$.

  v) Results of Model I in $P_4$ and $P_6$ approximations are
  in good agreement for $F_0$. For $F_1$ there are still some small
  differences.

  A few comments are as follows:

  a) Results in Model II and Model III converge faster than
  the results in Model I. This is in agreement with the
  observation that higher moments in Model II and Model III are
  more suppressed than those in Model I. Hence Model II and Model III
  should have better convergence properties.

  b) Value of $F_0$ at large radius can be understood using flux
  conservation. The flux of neutrino is given by
  \bea
  Tr[\rho_0-\rho_1]=z^{-2} Tr[\rho^\prime_0-z^{-2} \rho^\prime_1]
  \label{flux}
  \eea
  The flux, as it should be, scales as $z^{-2}$( or $r^{-2}$) in stationary
  approximation. So $F_0-z^{-2} F_1$ is a conserved quantity. At large $r$ this
  quantity approaches to $F_0$. On the other hand its initial value can be read out
  directly from the models of neutrino emission. Using
  Eqs. (\ref{model1}), (\ref{model2}) and (\ref{model3}) we find
  that at large $r$
  \bea
 && F_0 \to \frac{1}{2}, ~~\textrm{in Model I} \\
 && F_0 \to \frac{2}{3}, ~~\textrm{in Model II} \\
 && F_0 \to \frac{1}{2}, ~~\textrm{in Model III}
  \eea
 These values are in agreement with the plots in Figs. \ref{fig2},
 \ref{fig3} and \ref{fig4}.

 c) The scaling behavior of $F_1$ tells us that in $P_N$
 approximation with $N>1$ the self-interaction Hamiltonian
 scales down faster than $r_0^4/r^4$.

 d) Numerical study shows that $Tr[\rho^\prime_k]$ with $k>1$
 also drops down by $10^{-1}-10^{-2}$ at large $r$. It is
 a further support to the point that moment equations converge
 quite fast.

 \section{Flavor transformation}\label{sec4}

 In this section we do some analysis on flavor transformation of
 supernova neutrinos. We study the case of inverted mass hierarchy and
 for simplicity we neglect matter effect in the analysis.

      \begin{figure}
\begin{center}
\includegraphics[height=7.cm,width=16cm]{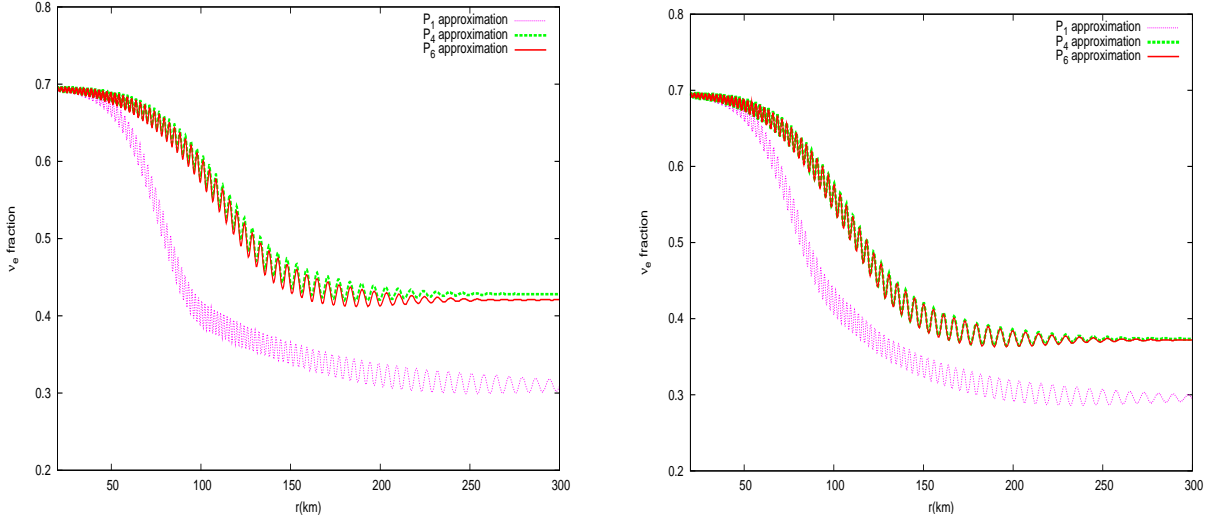}
\end{center}
 \vskip 0.0cm
  \caption{\small (color online) Fraction of $\nu_e$,
  $n_{\nu_e}/(n_{\nu_e}+n_{\nu_x})$, versus radius $r$ in in different
  models. Left in Model I; Right in Model II.
  $|\Delta m^2_{31}|=3 \times 10^{-3}$eV$^2$, $\sin^2 2\theta_{13}=0.01$.
   }
 \label{fig5}
\end{figure}

  In Fig. \ref{fig5} we give plots of $\nu_e$ fraction versus
  radius. These plot are obtained by solving Eq. (\ref{LVEq}) numerically.
  For a small step we get
 \bea
 \rho^\prime_k(r+\Delta r)=-\frac{\Delta r}{r_0} Q^1_k[\rho^\prime_l(r)]
 +e^{-i H_A \Delta r} \rho^\prime_k(r) e^{i H_A \Delta r}
 \label{solscheme}
 \eea

 In these plots one can see synchronized oscillation for which neutrinos
 of all energy point to the same direction in flavor space.
 Beyond the region of synchronized oscillation neutrino flavor vectors spin down
 which leads to neutrino flavor conversion.

 We compare numerical results of $P_4$ approximation and of $P_6$
 approximation in models I and III. We find nice agreements between these two
 approximations. This shows that $P_N$ approximation converge quite
 fast. This is in agreement with the discussion in the last section
 that the scaling law of the strength of the Hamiltonian converge quite
 fast.

 In Fig. \ref{fig5} one can see that result of $P_1$ approximation
 is quite different from that of $P_4$ and $P_6$ approximations.
 This is also consistent with discussion in the last section. Since
 the scaling law of the Hamiltonian in $P_1$ approximation is quite
 different from that in $P_{4,6}$ approximation we would expect to
 find difference in oscillation pattern.

     \begin{figure}
\begin{center}
\includegraphics[height=7.cm,width=10cm]{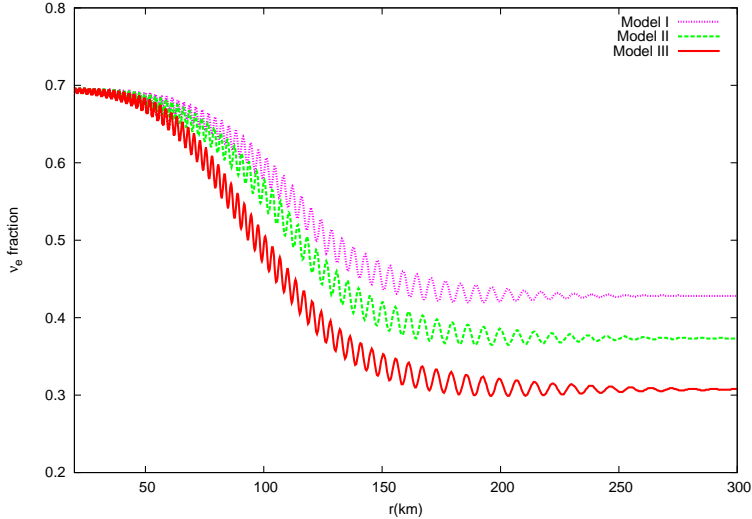}
\end{center}
 \vskip 0.0cm
  \caption{\small (color online) Fraction of $\nu_e$,
  $n_{\nu_e}/(n_{\nu_e}+n_{\nu_x})$, versus radius in different
  models. Neutrino parameters are the same as in Fig. \ref{fig5}.
   }
 \label{fig6}
\end{figure}

 In Fig. \ref{fig6} we compare numerical results in models I, II and
 III. One can see that there are some differences in the oscillation
 pattern.  This is consistent
 with the analysis on the strength of self-interaction
 Hamiltonian in these models. Numerical results show that at large
 $r$
 \bea
 && F_1 \to 0.126, ~~\textrm{in Model I} \\
 && F_1 \to 0.134, ~~\textrm{in Model II} \\
 && F_1 \to 0.151, ~~\textrm{in Model III}
 \eea
 Since differences in Hamiltonian are not large at large radius and
 the differences in the oscillation pattern should not be large either.

 We note that the scaling law of $F_1$, hence the amplitude of
 neutrino self-interaction Hamiltonian, is model dependent. This
 dependence on model is nicely described by the corrections given
 by higher moments in moment equations. Previous researches use
 fixed scaling function for the self-interaction
 Hamiltonian and do not take into account the dependence of the
 scaling law on the emission angle distribution of neutrinos. As
 a comparison one can check the result using a fixed scaling
 function. For example, one can use $\rho_1= S^2 \rho_1(r_0)$ where
 $S(r)=z^2/(1+\sqrt{1-z^2})$. Hence $F_1=0.5/(1+\sqrt{1-z^2})^2$
 in model I,$F_1=0.33/(1+\sqrt{1-z^2})^2$ in model II and
 $F_1=0.5/(1+\sqrt{1-z^2})^2$ in model III. At large radius
 this model independent scaling function gives
 $F_1\to 0.125$ in model I, $F_1\to 0.083$ in model II
 and $F_1\to 0.125$ in model III. Only for model I
 this fixed scaling function gives a correct result at large radius.
 In model II this fixed scaling function gives a value quite different from
 the value obtained using moment equations.

 \section{Conclusion}\label{sec5}
 In summary we have analyzed
 some properties of moment equations and the flavor transformation
 of supernova neutrinos. We have analyzed the scaling behavior
 of neutrino density and the amplitude of self-interaction
 Hamiltonian of neutrinos. They are related to quantities $\rho_{0}$ and
 $\rho_1$.

 We analyzed the convergence of $P_N$ approximation of moment equations.
 Numerical results show that the scaling behavior of $Tr[\rho_{0,1}]$
 converge for $N< 10$. We show that results
 of neutrino oscillation also converge quite fast. These analysis
 are consistent. Since the integration of $E^2 \rho_1$ give the
 self-interaction Hamiltonian the analysis on $Tr[\rho_1]$ tell
 us how fast the amplitude of neutrino self-interaction
 converge.

 We analyze neutrino flavor transformation. We find synchronized
 oscillation and bipolar oscillation in the oscillation pattern
 of supernova neutrinos. We find that
 oscillation pattern of neutrinos converge quite fast for
 $N<10$. The $P_1$ approximation can be used to make
 qualitative analysis but can not be used to do precise numerical
 study.

 We study three models of emission angle distribution
 of neutrinos on the oscillation pattern and analyze model
 dependence of neutrino flavor transformation
 on the emission angle distribution of neutrinos.
 Different models of emission angle distribution can give different
 results in the scaling
 behavior of self-interaction Hamiltonian and in oscillation pattern of
 neutrinos. This model dependence is carefully taken into account in
 the correction given by higher moments in moment equations.

 Previous works on oscillation of supernova neutrinos use fixed scaling
 function for the self-interaction Hamiltonian and do not
 take into the fact that the scaling law can be different in different
 models of neutrino emission. Analysis on the model
 dependent effect of emission angle distribution in neutrino oscillation
 is not presented in previous works.
 \\

 Acknowledgement:
 I wish to thank Y. Z. Qian, G. Raffelt, A. Yu. Smirnov  and
 H. Duan for
 discussions on neutrino flavor conversion in supernova.

\end{document}